\documentstyle[aps,preprint]{revtex}

\input{psfig.sty}

\newcommand{\ar}{\arrowvert}

\tighten

\begin{document}
\hyphenation{english}
\title{QCD Glueball Regge Trajectories and the Pomeron}
\author{Felipe J. Llanes-Estrada and Stephen R. Cotanch}
\address{Department of Physics, North Carolina State
University, Raleigh, NC 27695-8202}
\author{Pedro J. de A. Bicudo and J. Emilio  F. T. Ribeiro}
\address{Departamento de F\'{\i}sica and Centro de F\'{\i}sica das
Intera\c{c}\~oes
fundamentais,
Instituto Superior T\'ecnico, Av. Rovisco Pais, 1049-001 Lisboa, Portugal}
\author{Adam Szczepaniak}
\address{Department of Physics and Nuclear Theory Center, Indiana University,
Bloomington, IN 47405}

\date{\today}
\maketitle

\begin{abstract}
We report a glueball Regge trajectory emerging from diagonalizing a
confining  Coulomb gauge Hamiltonian  for
constituent gluons. Using
a BCS vacuum ansatz and gap equation, the dressed gluons acquire a
mass,  of order 800 $MeV$, providing the quasiparticle
degrees of freedom for a TDA glueball
formulation. The TDA
eigenstates for two constituent gluons have orbital, $L$, excitations with
a characteristic energy of 400 $MeV$ revealing a clear
Regge trajectory for  $\vec{J} = \vec{L}
+ \vec{S}$, where $S$ is the total (sum) gluon spin.
Significantly, the $S = 2$ glueball spectrum coincides with the Pomeron
given by $\alpha_P(t)=1.08+0.25 \ t $.
Finally, we also ascertain that lattice data
supports our result,  yielding an average intercept of 1.1 in good
agreement with
the Pomeron.
\end{abstract}

\vspace{1cm}
Pacs \#: 11.55.Jy;  
12.39.Mk; 
12.39.Pn 
12.40.Yx 
Keywords: Glueball Regge trajectories, Pomeron, QCD, Coulomb gauge Hamiltonian,
TDA.

\vspace{1cm}

\newpage

A vintage problem in hadronic physics is to fundamentally explain
the observed monotonously rising, to asymptotically flat cross section
behavior with
increasing  Mandelstam variable $s$. Pre-QCD cross section
theorems
\cite{theorems}  involving $t$ channel exchanges motivated
Pomeranchuk's
assertion \cite{pomeron}   that
vacuum quantum number ($J^{PC} = 0^{++}$) $t$ channel exchanges would
describe  elastic cross section high energy dependence. Related, it was
historically known that hadronic
resonances with specific quantum numbers could be connected to a series or
''tower'' of
excited states.  From this backdrop the successful Regge picture
\cite{Regge,Collins}, with poles and trajectories linear in
$t$ (mass squared), $\alpha(t) =
\alpha(0) + b \ t$, emerged which effectively described cross sections and
resonances and therefore unified both scattering and bound state
data. Of particular interest in this letter are analyses of high
energy  diffraction data
\cite{Donnachie,Pichowsky}
yielding a $J^{PC} = J^{++}$ Regge trajectory with the largest intercept
$\alpha(0)$. This trajectory is called the Pomeron and from detailed
fits is given by
\begin{eqnarray}
\alpha_P(t)=1.08+0.25 \ t \ .
\end{eqnarray}
Note that by definition, the Pomeron entirely governs and correctly
reproduces the
high energy scattering behavior since the cross section scales as
$s^{\alpha(0) -1}$.    The simplicity of this picture is very appealing and the
Pomeron remains a contemporary tool for understanding high energy
processes, both experimentally and theoretically
\cite{Thesis} .
In the discussion below we specialize to the soft non-perturbative
Pomeron (for an account of the BFKL hard Pomeron in perturbative QCD
see \cite{BFKL,Forshaw}).

\begin{figure}

\begin{picture}(280,390)(0,0)
\put(0,0){\psfig{figure=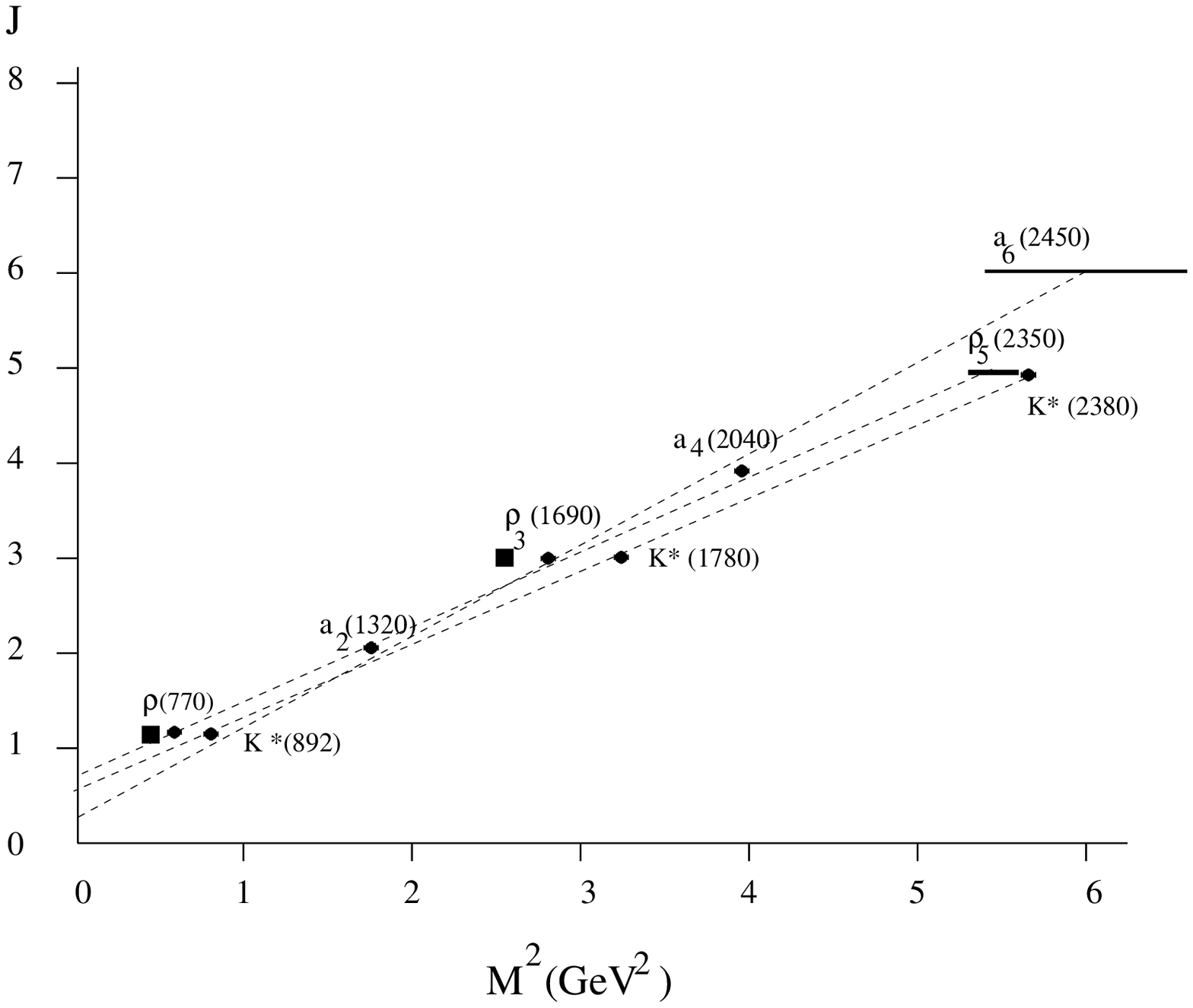}}
\end{picture}
\vspace{1cm}
\caption{Meson ($\rho$, $K^*$ and $a$) Regge trajectories constructed
from recent tabulated data (dark circles and error bars, PDG 2000). Boxes
are model TDA predictions for the $\rho$ trajectory.}
\label{Reggemeson}
\end{figure}

\begin{figure}

\begin{picture}(280,390)(0,0)
\put(0,0){\psfig{figure=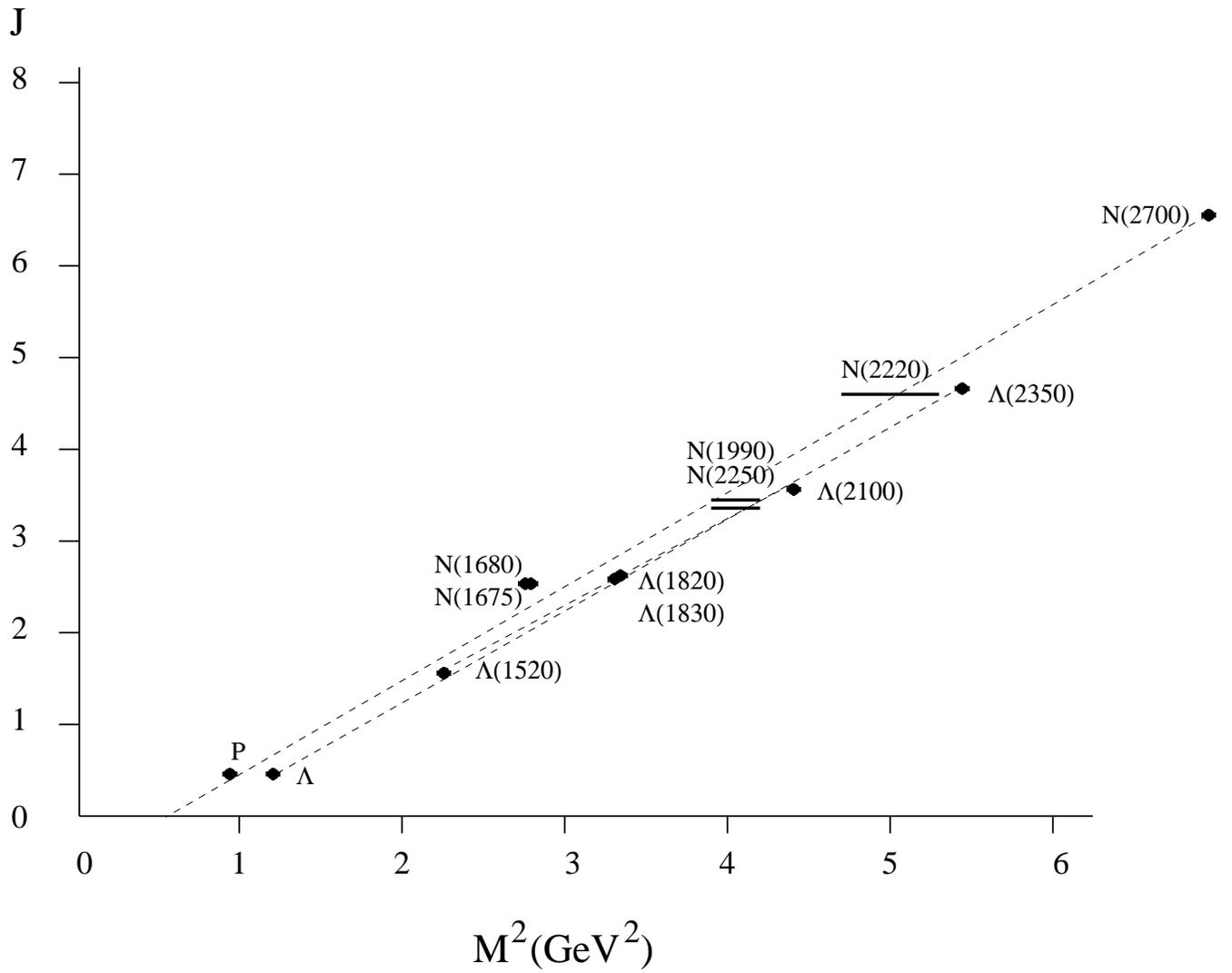}}
\end{picture}
\vspace{1cm}
\caption{Baryon ($N$ and $\Lambda$) Regge trajectories constructed
from recent tabulated data (dark circles and error bars, PDG 2000).}
\label{Reggebaryon}
\end{figure}

According to Regge theory, the trajectory describing scattering should
also be consistent with physical hadron states, implying
a $J^{PC} = J^{++}$ trajectory of mesons with slope near $0.25 \
GeV^{-2}$.
However, all  known hadronic resonances (both mesons and baryons)
fall on  Chew-Frautschi \cite{Chew} trajectories
with slopes close to $0.9 \ GeV^{-2}$ but different intercepts.
This is illustrated in Figs. \ref{Reggemeson} and \ref{Reggebaryon} for 
several mesons 
and baryons, respectively, 
including the $\rho$ whose negative parity, positive
G parity tower is well described by 
$\alpha _{\rho} (t) = .55  + 0.9 \ GeV^{-2} \  t$. 
This incompatibility led to the conjecture that
the Pomeron corresponded  instead to
a Regge trajectory for gluonic states, and most likely
one describing the $J^{++}$ glueballs.

Not surprisingly, the gluonic nature of the Pomeron
has been widely investigated  (see
\cite{Kisslinger} for a recent review)
with several recent studies linking the Pomeron and
glueball trajectories.  From meson trajectories and  Wilson loops,
Kaidalov and Simonov \cite{Kaidalov}
estimate that the glueball intercept
is $\alpha(0)\simeq .7 $, while
Soloviev \cite{Soloviev}
computes
$\alpha(0)= 1.07\pm 0.03 $
using the quantized elliptic Nambu-Gotto string model.
Motivated by the Maldacena conjecture,
and solving the dilaton wave equation,
Brower, Mathur and Tan \cite{brt}  obtained a
glueball intercept
of $\alpha(0)\simeq 1.2 $.
Finally, using mass relations between the glueball and meson sectors,
Brisudova, Burakovsky and Goldman \cite{Brisudova} obtain a glueball Regge
trajectory with slope $\simeq 0.3 \pm 0.1 \  GeV^2$.

In this communication we report results from a new constituent glueball
approach which further strengthens the gluonic Pomeron conjecture.
Similar to previous relativistic quark approaches \cite{johnson,nimai} yielding
a consistent meson Chew-Frautschi Regge slope,  our relativistic many-body
calculations also produce a glueball Regge trajectory that arises naturally
from the
confining linear potential and the scale of angular excitations.  In reproducing 
the lattice gauge positive parity glueball spectrum, we obtain a Regge
trajectory
with intercept near and above unity.  Most significantly, the trajectory
slope is close to the Pomeron value ($b \approx 0.25 \  GeV^2$) and
predominantly independent of
model details.
Our group has applied many body techniques to develop a unified approach
for the hadron  spectrum utilizing a QCD inspired Hamiltonian. Both
the  gluon \cite{ssjc96} and meson  \cite{flesrc,flsc} sectors
have been realistically described and
a hybrid meson calculation has also recently been completed
\cite{flschy}.

We briefly summarize our many-body approach (see refs.
\cite{ssjc96,flesrc,flsc}
for complete calculational details).
The starting point is a Coulomb gauge,
field theoretical Hamiltonian
\begin{eqnarray} \label{Hamiltonian}
H = Tr \int d {\bf x} ( {\bf \Pi}^a  \cdot  {\bf \Pi}^a +
{\bf B}_{A}^a  \cdot {\bf B}_{A}^a ) - \frac{1}{2} \int d {\bf x} d {\bf y}
\rho^a({\bf x})V(\arrowvert {\bf x} -
{\bf y} \arrowvert) \rho^a({\bf y}) \ ,
\end{eqnarray}
with color charge density $\rho^a = f^{abc} {\bf A}^b \cdot {\bf \Pi}^c$
and fields ${\bf \Pi}^a, {\bf B}_{A}^a =  \nabla \times 
{\bf A}^a $.  We utilize an instantaneous Cornell potential for confinement,
$V = -\frac{\alpha_s} {r} + \sigma r$.  
The potential parameters are
$\alpha_s = .2$,
and, from independent lattice results, $\sigma=\frac{3}{4} \ 0.18 \ GeV^2$. 
The gluons are then dressed by means of a  Bardeen, Cooper
and Schrieffer (BCS) ansatz for the vacuum ground state which, through a
variational
Hamitonian minimization,
generates the gap equation
\begin{equation}
\label{gluongap3d}
\omega_q^2 = q^2 - \frac{3}{4} \int
\frac{d{\bf
k}}{(2\pi)^3}
\hat{V}(\arrowvert {\bf q}
-{\bf k} \arrowvert )(1+ (\hat{{\bf k}}\cdot \hat{{\bf q}})^2) \left(
\frac{w_k^2 - w_q^2}
{w_k} \right) \ ,
\end{equation}
for the gluon self-energy, $\omega_q$,
containing the dynamical gluon  mass.
Our  potential in
momentum space is  $\hat{V}(\ar 
{\bf q}-{\bf k} \ar)
= -\frac{4\pi\alpha_s}{({q}-{k})^2}
-\frac{8\pi\sigma}{({q}-{k})^4}$, where each term generates a 
divergence in the gap equation (Coulomb is quadratic, confinement is
logarithmic).  Since we are focusing on confinement we have
omitted the Couloub potential component and then 
used a cut-off parameter $\Lambda=4-5 \  GeV$ to regularize the
remaining logarithmic divergence.
Finally, a Tamm-Dancoff
(TDA) diagonalization, truncated to the 1p-1h level, is performed for the
same field
theoretical Hamiltonian (no divergences, Coulomb potential is now included)
to generate the glueball
spectrum. We used the same linear potential 
in our quark sector applications which produced reasonable
results
\cite{flesrc,flsc}, including the appropriate meson Regge slope as
indicated by the solid squares in Fig. \ref{Reggemeson} for the $1^{--}$ and $3^{--}$ 
$\rho$ trajectory.  It is important to stress that our quark sector
application properly implemented chiral symmetry.

\begin{figure}
\psfig{figure=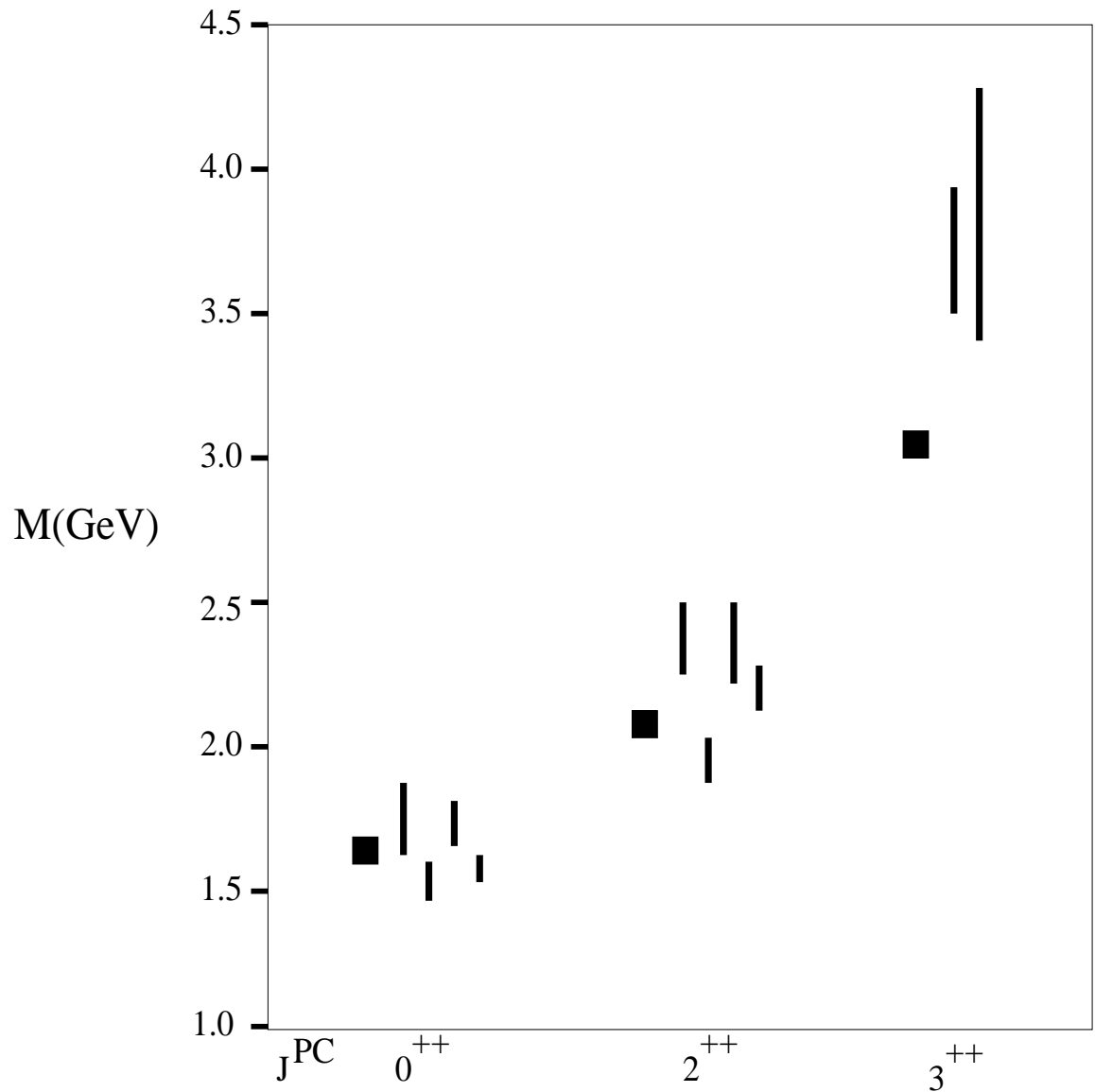,width=6in,height=6in}
\vspace{1cm}
\caption{The TDA (boxes) and lattice gauge (vertical bars showing error) 
glueball spectra.  The lattice results, from left to right,
are from refs. \protect\cite {mp,l1,l2,l3}, respectively.}
\label{spectrum}
\end{figure}

At the more technical level, the glueball TDA state, 
for fixed orbital, $L$, and total gluon spin, $S$,
is represented by
\begin{eqnarray}
|\Psi^{JPC}_{LS}\rangle= \sum_{a m_1 m_2}\int{d{\bf
q}\over(2\pi)^3}\Phi^{JPC}_{LSm_1
m_2}({\bf q}) \alpha_{m_1}^{a\dagger}({\bf q})\alpha_{m_2}^{a\dagger}(-{\bf
q})|\Omega\rangle
\,,  \label{eq:3.6}\end{eqnarray}
where the $\alpha_{m}^{a\dagger}$ are quasiparticle creation operators
acting on the
BCS vacuum state $| \Omega \rangle$. The sum is over
the color index
$a$ and the two
transverse spin projections,
$m = 1, 2$.  The 
glueball wavefunction has angular momentum composition  
\begin{eqnarray}
\Phi^{JPC}_{LSm_1 m_2}({\bf q}) = \langle 1 m_1 1 m_2 \arrowvert S
m_S \rangle  \langle L m_L S m_S \arrowvert J m_J \rangle
\phi^{JPC}_{LS}(q) Y_L^{m_L}(\hat {\bf { q}}) \ .
\end{eqnarray}
As mentioned above, the gluon quasiparticle energies are obtained
from the nonlinear mass gap
equation which yields a gluon mass around 800 $MeV$. Then, 
with the gap energy, the
linear TDA equations are diagonalized for the $J^{PC}$ states of interest,
either as
a radial equation or variationally in a multi-dimensional Monte Carlo
calculation for the  Hamiltonian matrix
elements using the computer code VEGAS. Our work omits renormalization,
but a recently improved, renormalized glueball calculation
\cite{gjc} produced similar $J^{PC} = 0^{\pm+}$ results.  Using a new, independent
code we have recomputed  
the glueball spectrum and confirmed our original results \cite{ssjc96}.
We have also extended the analysis to the maximally acceptable 1p-1h
excitation range spanning higher angular momentum states,
up to $3^{++}$.  The 1p-1h energy upper bound is
about  3 $GeV$ since it is well known that 
2p-2h (4 gluon)
states become important for energies twice the lightest
1p-1h excitation ($\approx $ 1.6 $GeV$ ground state glueball).  
A realistic calculation above 3 $GeV$  would entail mixing with
2p-2h states requiring a complicated (4 gluon) Fock space diagonalization.  
Further,  higher
angular momentum
significantly enhances the interaction spin dependence which is
not accurately  known.  These two reasons are also why we only predict 
states up to $J = 3$ for the meson 
spectrum shown in Fig. \ref{Reggemeson}.    
The positive parity glueball states reliably
predicted by our model are compared to the
previous and most recent lattice measurements \cite{mp} in Fig. \ref{spectrum}.

\begin{figure}
\psfig{figure=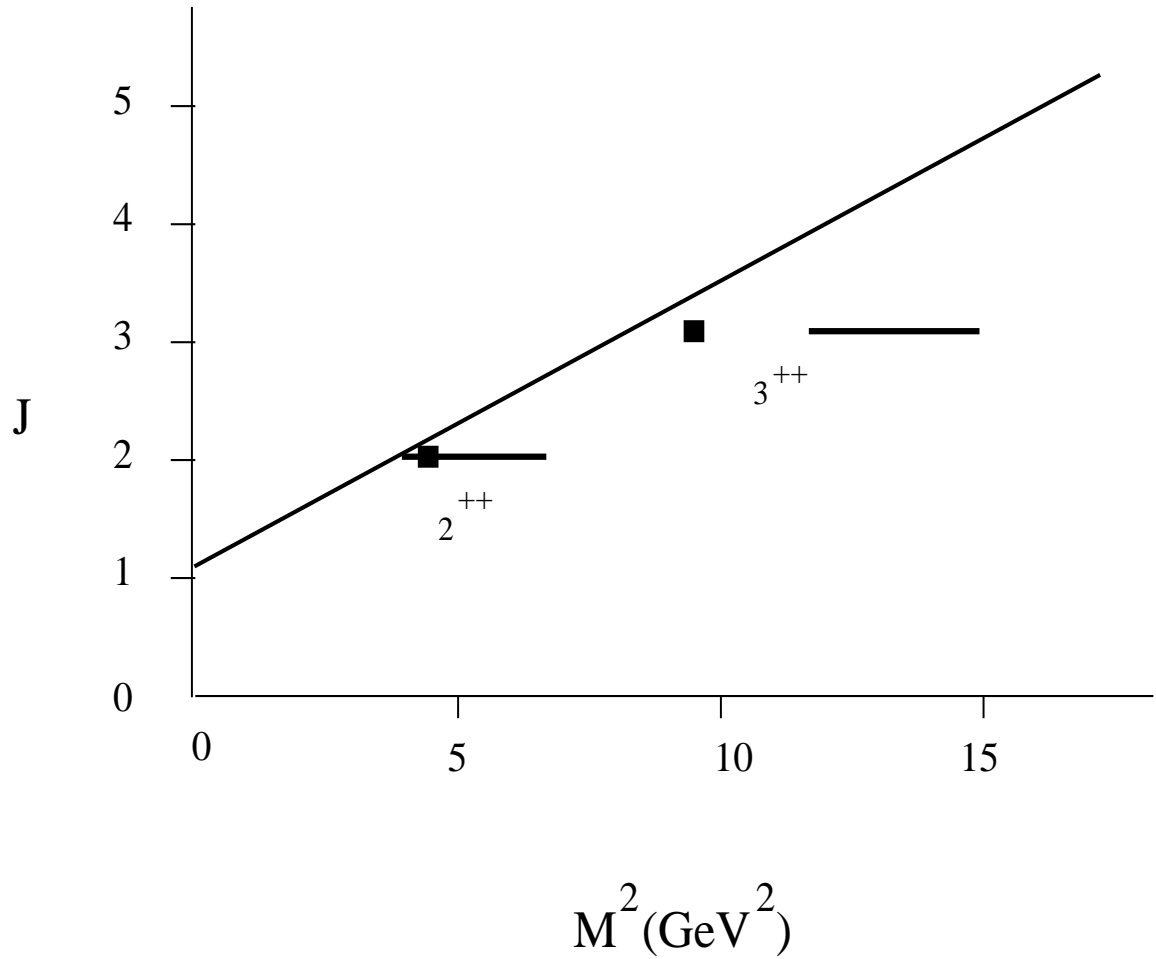,width=6in,height=5in}
\vspace{1cm}
\caption{Comparison of TDA (boxes) and lattice (horizontal bars)
$J^{++}$ glueballs with the Pomeron (solid line).}
\label{Reggelattice}
\end{figure}

Wavefunction symmetry for two identical bosons requires that $L+S$ must be
even. Since  the Pomeron should carry positive parity and
C-parity, L odd
is not considered nor three gluon states which have odd C-parity.
Also, according to Yang's theorem, a combination of
gluon transversality (Coulomb gauge) and Bose statistics, $J = 1$ states 
are forbidden
for a two gluon system.  Further, the $J=0$ state is precluded from
the Pomeron since it necessarily belongs to a lower (daughter)
trajectory. Therefore only the $J = 2^{++}$, with $L =0, S =2$,
and
$J = 3^{++}$, with $L =2, S =2$, states in Fig. \ref{spectrum} are Pomeron  
candidates and
the corresponding glueball Chew-Frautschi diagram is shown in 
Fig. \ref{Reggelattice} along
with the Pomeron trajectory.

Figure \ref{Reggelattice} constitutes the key findings of this study.  Notice that our
model yields a trajectory with slope .20 $GeV^{-2}$ and intercept 1.1, in 
reasonable agreement with the Pomeron. We found
little
sensitivity of our results to the confining potential parameter
as
increasing the strength by 33 \% (i.e. $\sigma = \frac{3} {4} 0.18 \
GeV^2
\rightarrow
0.18 \ GeV^2$), only marginally reduced the intercept. 
Further, and of interest to BFKL PQCD studies,
our model predicts the soft Pomeron is a correlated two gluon state with 
total intrinic
spin $S = 2$.   

Equally significant are the lattice
data which qualitatively supports our main result.  The lattice calculations
entail some uncertainty and therefore generate a less accurate trajectory
having slope .15 $\pm$ .05
$GeV^{-2}$.
However, the lattice intercept is above unity,
although
again with appreciable error (1.1 $\pm$ .5).  We therefore conclude 
that both our model and recent lattice data
are consistent with a glueball interpretation of the Pomeron.  

Ideally a $J=4^{++}$ (and higher) state should also be predicted. This may
be possible for lattice calculations but, as discussed above, would be 
quite formidable in our
approach.   
Conversely, assuming a glueball Pomeron 
interpretation
we can use the fitted trajectory to predict that the $4^{++}$ and $5^{++}$ glueballs should have
masses $M_J = 2 \sqrt{J - 1.08} = $ 3.42 and 3.96 $GeV$, respectively.

Finally, we offer a qualitative explanation for our unified
model
predictions of
meson and Pomeron slope trajectories with respective
values $b_{q \overline
{q}}
\approx 0.9
\ GeV^2$ and $b_{gg}\approx 0.25 \
GeV^2$.  In a simple QCD  string model with
massless partons separated by a
string with length $r$ and tension (or
energy per unit
length)
$k$, the
total relativistic mass scales as $kr$ while the angular momentum
is
proportional to $\frac{1} {2} k r^2$,  which generates a Regge
trajectory
with slope
$b = \frac{1} {2 \pi k r}$.  This string feature is
incorporated in our
model through
the linear confining potential ($\sigma
r$), yielding effective string
tensions,
$k_{q \overline {q}} = C_{q
\overline {q}} \sigma $ and
$k_{gg} = C_{gg} \sigma $, for the meson and
glueball systems,
respectively. Here
$C_{q \overline {q}} = \frac{N^2_c -1}
{2N_c}$ and $C_{q \overline {q}} =N_c$
are the respective  quark and gluon
color Casimir operators from the
density-density
Hamiltonian kernel. Hence for $N_c = 3$ the
meson and glueball
trajectory slopes are related by $b_{gg} =
\frac{C_{q
\overline
{q}}}{C_{gg}} b_{q \overline {q}} = \frac{4} {9} b_{q
\overline {q}} $.  While this simple estimate explains much of the
meson/glueball slope difference, it is
significant that our model is able
to account for entirely all of the
difference,
the rest due to BCS
quasiparticle masses and the Coulomb potential in the TDA.  It
would be
very
interesting to see if alternative QCD models, particularly the flux
tube, can
reproduce these results.

In summary, our
relativistic many-body approach provides a glueball Regge trajectory 
similar to the Pomeron diffraction fit. The model also unifies and describes
the quark sector by properly implementing chiral symmetry (and breaking) to
reproduce the observed  meson spectra and attending  Regge
trajectories
with the same, predetermined confining potential. Recent
lattice measurements are also consistent with this
picture and support a gluonic interpretation of the Pomeron.  Reproducing
higher $J^{++}$ states on the Pomeron trajectory would confirm this 
assertion and  more sophisticated model
calculations are
in progress along with a  call for alternative model predictions.
Finally, our results indicate that the Pomeron provides a worthwhile guiding
constraint which future theoretical and experimental glueball studies
should utilize. 

Pedro Bicudo acknowledges enlightening Pomeron discussions
with Barbara Clerbaux, Brian Cox and Mike Pichowsky.
This work is supported in part by grants DOE DE-FG02-97ER41048,
DE-FG02-87ER40365 and NSF INT-9807009.
Felipe J. Llanes-Estrada was a SURA-JLab graduate fellowship
recipient and thanks Katja Waidelich for technical help. Supercomputer time
from NERSC is also acknowledged.


\newpage

\end{document}